# The study of electronic nematicity in an overdoped $(Bi, Pb)_2Sr_2CuO_{6+\delta}$ superconductor using scanning tunneling spectroscopy

Yuan Zheng[1], Ying Fei[1], Kunliang Bu[1], Wenhao Zhang[1], Ying Ding[2], Xingjiang Zhou[2,3], Jennifer E. Hoffman[4], Yi Yin[1,5,*]

1. Department of P23hysics, Zhejiang University, Hangzhou, 310027, China.
2. Beijing National Laboratory for Condensed Matter Physics, Institute of Physics, Academy of Science, Beijing 100190, China.
3. Collaborative Innovation Center of Quantum Matter, Beijing 100871, China.
4. Department of Physics, Harvard University, 17 Oxford St. Cambridge, 02138, USA.
5. Collaborative Innovation Center of Advanced Microstructures, Nanjing 210093, China.
   *Correspondence and requests for materials should be addressed to Y. Y. (email: yiyin@zju.edu.cn).

**Abstract**

The pseudogap (PG) state and its related intra-unit-cell symmetry breaking remain the focus in the research of cuprate superconductors. Although the nematicity has been studied in $Bi_2Sr_2CaCu_2O_{8+\delta}$, especially underdoped samples, its behavior in other cuprates and different doping regions is still unclear. Here we apply a scanning tunneling microscope to explore an overdoped $(Bi, Pb)_2Sr_2CuO_{6+\delta}$ with a large Fermi surface (FS). The establishment of a nematic order and its real-space distribution is visualized as the energy scale approaches the PG.

The electronic nematic order has attracted a lot of interests in both the cuprate and iron-based high-transition-temperature (high-$T_c$) superconductors [1-3]. In solid state systems with the nematic order, the rotational symmetry is broken while the translational lattice symmetry is preserved [4]. Various experimental techniques, such as x-ray scattering, neutron scattering, angle-resolved photoemission spectroscopy (ARPES), and scanning tunneling microscope (STM), have been applied to determine the nematicity through the evidences of symmetry breaking in orbital and magnetic orders [5-8]. The understanding about nematic order may help unravel the mechanism of the superconducting (SC) state [9-11].

STM is a powerful tool for detecting electronic structures with an atomic resolution in the real space. For $Bi_2Sr_2CaCu_2O_{8+\delta}$ (Bi-2212), the STM studies have revealed a 90° rotational symmetry breaking for the O sites within each $CuO_2$ unit cell, around the energy level of the pseudogap (PG) states [4]. The nematicity is suggested to be a relevant order parameter of the PG state which strongly intertwines with the SC state. The STM studies of Bi-2212 have further clarified that the nematic order in general decreases with doping and finally disappears when a small-to-large Fermi surface (FS) reconstruction occurs at carrier doping of $p \sim 0.19$ [12].

With similar phase diagrams, various cuprate superconductors can differ in the detailed electronic structures [13-14]. As the doping increases, the small-to-large FS transition also occurs in (Bi, Pb)$_2$Sr$_2$CuO$_{6+\delta}$ (Bi-2201) at $p \sim 0.15$. However, the PG phase extends to the overdoped regime for Bi-2201 [14-15] while terminating in the SC dome for Bi-2212. It is thus physically interesting to investigate the nematic order in overdoped Bi-2201 where the PG state coexists with a large FS. The study in this regime may enable us to explore the early stage in the formation of the nematicity. In this paper, we apply STM to study an overdoped Bi-2201 sample through an analysis approach similar to that in Ref. [4]. Our studies confirm the establishment of the nematic order and reveal a strong real-space fluctuation of site-specified order parameters.

**Results**
**Topographical and electronic properties of overdoped Bi-2201.** The samples studied in this paper are overdoped (Bi, Pb)$_2$Sr$_2$CuO$_{6+\delta}$ single crystals with $T_c = 13$ K and all the data are taken in an ultra-high-vacuum STM at $T = 4.5$ K. Figure 1a displays a topographic image obtained on a cleaved BiO surface, showing a clear square lattice of Bi atoms with interatomic spacing of $a_0 \approx 3.8$ Å. A part of Bi atoms are substituted by Pb atoms, represented by brighter spots in the lattice, and the incommensurate supermodulation is completely suppressed due to the elimination of the periodic potential of strain [16].

The local electronic property can be probed by the differential conductance ($dI/dV$) spectroscopy, which is proportional to the local density of states (LDOS). A spectral survey is taken simultaneously with the topographic image in Fig. 1a, at a dense array of locations. All the following data are shown and analyzed in the same field of view (FOV) and the atomic registry is precisely maintained. Figure 1c presents a typical slice of the spectroscopy image or a differential conductance map, $g(\vec{r}, E)$, at a relative low bias voltage $V = 7.5$ mV. The wavelike spatial pattern of the LDOS modulation mainly originates from the Bogoliubov quasi-particle interference (QPI) [17-18]. In the Fourier transformed map, $\tilde{g}(\vec{q}, E = 7.5 \text{ meV})$ (see Fig. 1d), the dominant QPI wave vectors extracted around strong signals could provide the Fermi surface (FS) information. We observe a 'triplet' signal in the anti-nodal region and a trace extending from the nodal to anti-nodal regions. These two features are assigned as the signature of a large FS [14]. As a comparison, an 'octet' QPI associated with a small FS has been observed for underdoped Bi-2212 and Bi-2201 samples (see Fig. 1b) [13, 14, 17].

The Bogoliubov quasi-particle excitations, which is related with the SC phase, are often detected around the Fermi energy. The electronic excitations behave differently as the energy increases to the PG energy scale [13]. In Fig. 1e, we present another $g(\vec{r}, E)$ at a relative high bias voltage $V = 25$ mV. The wavelike pattern here is not from the QPI, but related with a quasi-localized charge order (also called the smectic order) [4, 9]. In the Fourier transformed map, $\tilde{g}(\vec{q}, E = 25 \text{ meV})$ (see Fig. 1f), the dominant wave vectors are near $\vec{q}^* \sim \pm \frac{3}{4}\vec{Q}_x$ and $\pm \frac{3}{4}\vec{Q}_y$, corresponding to the real-space periodicity of charge order modulation. Here $\pm \vec{Q}_x = \left(\pm \frac{2\pi}{a_0}, 0\right)$ and $\pm \vec{Q}_y = \left(0, \pm \frac{2\pi}{a_0}\right)$ are positions of Bragg peaks. Although both the intra-unit-cell nematicity and the charge order are developed when the energy approaches the PG magnitude (see Fig. 1b), these two orders are prominent at different wave vectors in the Fourier transformed map. The former is around the Bragg peaks, while the latter is around $\vec{q}^*$ [4, 9].

**Collective nematic order.** Although the current Bi-2201 sample is overdoped with a large FS, the PG state is known to exist [14, 15]. The PG magnitude ($\Delta_{\text{PG}}$) at each location in the FOV is determined by extracting the bias voltage of the positive coherence peak from its $dI/dV$ spectroscopy. The resulting PG map in Fig. 2a shows a strong nanoscale inhomogeneity where large-$\Delta_{\text{PG}}$ regions are spread in space and surrounded by low-$\Delta_{\text{PG}}$ regions. In the broad distribution of $\Delta_{\text{PG}}$, we also find ~10.9% zero gap patches, which are attributed to van Hove singularities (VHS, with details in Supplementary Note S1 and Fig. S1). The similar VHS behavior was specified in pure non-cation-doped Bi-2201 samples [19]. To detect the electronic order hidden by the strong inhomogeneity of the PG distribution, we apply a ratio map, $Z(\vec{r}, \varepsilon) = g(\vec{r}, \varepsilon)/g(\vec{r}, -\varepsilon)$ [20]. The reduced energy $\varepsilon = |E|/\Delta_{\text{PG}}(\vec{r})$ is rescaled with respect to the PG magnitude at each location. The ratio $Z(\vec{r}, \varepsilon)$ between the LDOS at two opposite reduced energies ($\pm \varepsilon$) can minimize the systematic error of $g(\vec{r}, \pm \varepsilon)$ caused by the setpoint effect [20].

Figure 2b displays a typical image of $Z(\vec{r}, \varepsilon = 0.975)$ in the same atomically resolved area as in Fig. 1a. The spatial average of $Z(\vec{r}, \varepsilon)$ is artificially assigned for the VHS regions (see Supplementary Note S1 and Fig. S1), which however does not affect the main conclusion of this paper. The electronic spatial patterns consist of an apparent charge order modulation and an underlying intra-unit-cell order. To obtain an atomic registry with picometer-scale precision, we implement a lattice drift correction of the topographic image in all the $dI/dV$ maps simultaneously acquired [4,

21]. After the lattice drift correction, we implement the Fourier transform and obtain a ratio map, $\tilde{Z}(\vec{q}, \varepsilon = 0.975)$, in the momentum space (Fig. 2c). Four sharp Bragg peaks at $\pm \vec{Q}_x$ and $\pm \vec{Q}_y$ are observed, each collapsing into a single pixel due to the drift correction. However, the two sets of Bragg peaks are not degenerate as revealed by the difference between $\mathrm{Re}\,\tilde{Z}(\pm \vec{Q}_x, \varepsilon = 0.975)$ and $\mathrm{Re}\,\tilde{Z}(\pm \vec{Q}_y, \varepsilon = 0.975)$ (shown in the inset of Fig. 2c). As a Cu site is selected as the origin when performing the Fourier transform, this difference signifies a symmetry breaking from $C_{4v}$ (90° rotational symmetry for four O sites surrounding each Cu site) to $C_{2v}$ (180° rotational symmetry for two O sites along the $x/y$ direction). A collective nematic order for this FOV of overdoped Bi-2201 is thus determined around the PG energy scale. The same investigation is applied to other reduced energies, which leads to the definition of a normalized order parameter,

$$\tilde{O}(\varepsilon) = [\mathrm{Re}\,\tilde{Z}(\vec{Q}_x, \varepsilon) - \mathrm{Re}\,\tilde{Z}(\vec{Q}_y, \varepsilon)]/\bar{Z}(\varepsilon), \qquad (1)$$

where $\bar{Z}(\varepsilon)$ is the spatial average of $Z(\vec{r}, \varepsilon)$. In Fig. 2d, the order parameter $\tilde{O}(\varepsilon)$ increases monotonically as $\varepsilon$ approaches the PG energy, illustrating a gradual establishment of the nematic order.

The information extracted from the Bragg peaks in the momentum space represents an asymmetric intra-unit-cell electronic modulation along the $x$ and $y$ directions in the real space. Based on the spatial structure shown in the inset of Fig. 3a, the dominant order in $\mathrm{Re}\,\tilde{Z}(\pm \vec{Q}_x, \varepsilon)$ and $\mathrm{Re}\,\tilde{Z}(\pm \vec{Q}_y, \varepsilon)$ arises from the O sites. In the real space ratio map of $Z(\vec{r}, \varepsilon)$, we take a summation over the unit cells and calculate an alternative order parameter [4],

$$O(\varepsilon) = \sum_n O(n, \varepsilon)/N = \sum_n [Z_y(\vec{r}_n, \varepsilon) - Z_x(\vec{r}_n, \varepsilon)]/[N\bar{Z}(\varepsilon)], \qquad (2)$$

where $\vec{r}_n$ is the location of the Cu site centered in the $n$th unit cell, and $N$ is the total number of unit cells. The results of $Z_x(\vec{r}_n, \varepsilon)$ and $Z_y(\vec{r}_n, \varepsilon)$ are averaged from the O sites of the $n$th unit cell along the $x$ and $y$ directions (labeled by $O_x$ and $O_y$), respectively. As the reduced energy $\varepsilon$ increases, $O(\varepsilon)$ follows the same trend as $\tilde{O}(\varepsilon)$, and a smooth transition from $C_{4v}$ to $C_{2v}$ is observed (as shown in Fig. 3a). The nematicity is a manifest from the asymmetric electronic structures on the $O_x$ and $O_y$ sites. As a comparison, we assign two set of copper sites, $Cu_1$ and $Cu_2$, along the two perpendicular directions, and the symmetry breaking of Cu sites is not found over the whole range of $\varepsilon$.

The nematicity defined in (2) is explored in the $Z(\vec{r}, \varepsilon)$ maps which are derived from the $dI/dV$ spectra. To confirm that the nematicity is not artificially induced by our data analysis, we directly average the $dI/dV$ spectrum over all the $O_x/O_y$ sites in the FOV. As shown in Fig. 3b, the $O_x/O_y$-averaged spectra are distinguished from each other in the regime of negative bias voltages. The dataset analyzed above was taken with a tunnel junction of sample bias $V_b = +100$ mV. Since the spectral weight is normalized for empty states (positive bias voltages), the spectral shift from asymmetric nematic order appears in the filled states (negative bias voltages). Under the condition of

$V_b = -100$ mV, we find the spectral shift of the $O_x$/$O_y$-averaged $dI/dV$ spectra in the empty states (see Supplementary Information Note S3 and Fig. S4). The symmetry breaking of $O_x$ and $O_y$ sites is a result of the asymmetric LDOS while the ratio $Z$-map provides a clearer identification of the nematic order parameter. Instead, the averaged $dI/dV$ spectra from $Cu_1$ and $Cu_2$ sites are indistinguishable over the whole energy range, which is consistent with the zero nematicity of Cu sites in Fig. 3a.

**Real-space distribution of the nematic order.** The two order parameters, $\tilde{O}(\varepsilon)$ and $O(\varepsilon)$, reflect the same phenomenon of the nematic order, despite the fact that they are defined separately in the momentum and real spaces. The real-space $Z(\vec{r}, \varepsilon)$ map will be further applied to explore the spatial distribution of nematicity. However, the electronic modulations other than the intra-unit-cell periodicity lead to strong interference signal in the real space, and the atomic-scale nematic order is buried in background noise and other electronic orders. A solution to this difficulty is to retain $\tilde{Z}(\vec{q}, \varepsilon)$ within a limited region around four Bragg peaks in the momentum space, which can be realized by

$$\tilde{Z}_f(\vec{q}, \varepsilon) = \tilde{Z}(\vec{q}, \varepsilon)[f_\Lambda(\vec{q} + \vec{Q}_x) + f_\Lambda(\vec{q} - \vec{Q}_x) + f_\Lambda(\vec{q} + \vec{Q}_y) + f_\Lambda(\vec{q} - \vec{Q}_y)]. \qquad (3)$$

The Gaussian filtering function, $f_\Lambda(\vec{q}) = \exp(-q^2/2\Lambda^2)$, is defined with a filtering size $\Lambda^{-1}$. Figure 3c presents a typical result of $\tilde{Z}_f(\vec{q}, \varepsilon = 0.975)$ filtered from Fig. 2c using $\Lambda^{-1} = 1.05$ nm. The inverse Fourier transform is subsequently applied to $\tilde{Z}_f(\vec{q}, \varepsilon = 0.975)$ for a real-space map of $Z_f(\vec{r}, \varepsilon = 0.975)$, as shown in Fig. 3d. Compared to the original map in Fig. 2b, this filtered map reduces longer-wavelength modulations while the spatial inhomogeneity is partially preserved. In the real space, the filtration is equivalent to a locally weighted average for each spatial location so that the background signal can be significantly suppressed. Following the definition in equation (2), we calculate the order parameter $O_f(\varepsilon)$ after the filtration, and the comparison with $O(\varepsilon)$ in Fig. 2a reveals the same development of the nematicity. Although the measurements of the collective order, $O_f(\varepsilon)$ and $O(\varepsilon)$, are similar, the filtering procedure is important for detecting the spatial distribution of nematic order. For each $n$th unit cell, the difference between the electronic structures of the O sites leads to the order parameter of this cell,

$$O_f(n, \varepsilon) = [Z_{f,y}(\vec{r}_n, \varepsilon) - Z_{f,x}(\vec{r}_n, \varepsilon)]/\bar{Z}_f(\varepsilon), \qquad (4)$$

where $Z_{f,x}(\vec{r}_n, \varepsilon)$ and $Z_{f,y}(\vec{r}_n, \varepsilon)$ are averaged over the O sites along the $x$ and $y$ directions. A coarse-grained average within the area of $\Lambda^{-2}$ is implied in our cell order parameter, which is consistent with a previous approach of estimating the correlation length by a local Fourier transform [4]. Consequently, we construct a new map of the cell nematic order parameter $O_f(n, \varepsilon)$. The image resolution of $O_f(n, \varepsilon)$ is reduced compared to that of $Z_f(\vec{r}, \varepsilon)$, as the information of each unit cell is compressed into a single pixel.

Three typical cell order parameter maps with $\varepsilon = 0.150, 0.375,$ and $0.975$ are displayed in Fig. 4a-4c, to demonstrate the evolution of $O_f(n, \varepsilon)$ with the change of the reduced energy, while the full evolution is provided in Supplementary Fig. S3. At each reduced energy, the order parameter map is inhomogeneously distributed. Both positive and negative nematicities form nanoscale domains, which are depicted in red and blue colors in Fig. 4a-4c. For a small reduced energy ($\varepsilon = 0.150$, in Fig. 4a), these two opposite nematic domains occupy roughly the same percentage of the image, leading to nearly zero collective nematicity. With the increase of the reduced energy, the colors of both domains are intensified so that the local nematicity is enhanced. In our selected FOV, the average strength of the positive nematicity however experiences a stronger enhancement than that of the negative nematicity. At the same time, the size of red domains grows, showing a gradual formation of the collective positive nematicity. For example, a blue domain at the top left corner of Fig. 4a is cut into two parts and a red domain emerges in Fig. 4b when the reduced energy changes from 0.150 to 0.375. On the other hand, the major domain structure is considerably preserved. With the Ising symmetry $x_n = \pm 1$ assigned to a red or blue pixel for each unit cell, a correlation function, $C = \sum_n |x_n - x'_n| / \sum_n (|x_n| + |x'_n|)$, is used to quantitatively describe the structural difference between the images of local order parameters at two different reduced energies. A similarity over 75% is found between Fig. 4a and 4b. As the reduced energy further approaches the PG energy level ($\varepsilon = 0.975$, in Fig. 4c), both the size increase of red domains and the color intensification of the two domains slow down, since the cell and collective nematic orders start to be saturated. The electronic structure in overdoped Bi-2201 follows a smooth transition of symmetry breaking, and a nanoscale disorder is frozen during the establishment of the nematicity.

To gain a more quantitative description of the site-specified nematicity, we draw a histogram of $O_f(n, \varepsilon)$ for each reduced energy $\varepsilon$ by counting $O_f(n, \varepsilon)$ over a selected bin size $\delta O_f = 0.001$. In Fig. 4e, the histograms resulted from Fig. 4a-4c are plotted, each being well fitted by a Gaussian envelope, $F(O_f) \propto \exp[-(O_f - O_f^G)^2 / 2\sigma_f^2]$, with a mean $O_f^G(\varepsilon)$ and a standard deviation $\sigma_f$. The same good accuracy of Gaussian fitting is found for all the other histograms. The center of each Gaussian distribution, $O_f^G(\varepsilon)$, provides another estimation of the averaged collective nematic order. As shown in Fig. 4f, the result of $O_f^G(\varepsilon)$ is almost the same as the collective order parameter $O_f(\varepsilon)$ defined in equation (2) after considering the error of the bin size. The positive collective nematic order in the selected FOV is formed by the shift of $O_f^G(\varepsilon)$ to the right, consistent with the gradual growth of the red domains in Fig. 4a-4c. Due to its strong inhomogeneous nature, $O_f(n, \varepsilon)$ is always broadly distributed with a large standard deviation due to the two types of nematic domains in nanoscale. The value of $\sigma_f$ in general increases with $\varepsilon$, as demonstrated by the gradual color intensification in Fig. 4a-4c. The evolution of $O_f^G(\varepsilon)$ and $\sigma_f(\varepsilon)$ suggests that the nematic order is first randomly generated by a real-space fluctuation and then is enhanced following the increase of the reduced energy. In the development of a positive nematicity extended over a larger region in the FOV, some blue domains are sustained possibly due to a non-negligible energy cost for flipping microscopic nematic orders. The broadening of the Gaussian distribution is however less severe than the shift of its center position, with the relative ratio of $\sigma_f(\varepsilon)/O_f^G(\varepsilon)$ consistently reduced as $\varepsilon$ increases (see the inset of Fig. 4f).

**Discussion**

The STM experiment reported in this paper confirms the nematic order in overdoped Bi-2201,

extending previous studies in Bi-2212 [4]. The collective nematic orders with rotational symmetry breaking at O sites are observed in both momentum and real spaces. A filtering procedure allows us to extract the site-specified intra-unit-cell signal and reveal the spatial distribution of electronic nematicity. In an underdoped Bi-2212 with a strong nematicity, the dominant order can extend over a few tens of nanometers [4]. In our overdoped Bi-2201 with a weak nematicity, the two opposite orders coexist when a nonzero collective order is developed around PG. Nanoscale domains are aggregated in real space for each of two opposite nematicities, which are induced by local fluctuations. With the increase of the reduced energy, the domain size of one nematicity increases, and its order strength gradually dominates although both orders are enhanced. The collective nematic order is established by the combination of these two effects. The evolution of the nematicity in our sample can help build a multi-step picture for the order formation.

In Bi-2212, the nematic order is determined in underdoped and optimal samples, but disappears for overdoped samples without the PG state. Our observation in the overdoped Bi-2201 presents the nematicity in cuprates with a large FS, implying the nematicity more correlated with the PG state than the FS structure. On the other hand, the nematic order also coexists with other electronic orders, e.g., the charge order (smectic order) which is interpreted by a Landau-Ginzberg theory [9]. The charge order is recently identified as a $d$-form factor order with a sophisticated phase sensitive analysis [22-24]. The PG state and related charge order are believed to originate from strong-correlated doped Mott insulator [25-26], whereas the large FS is often related with a Fermi liquid behavior [2]. Future studies will be necessary to further explore the connections between the nematicity, charge order and FS.

## Methods

**Sample growth.** The high-quality Pb-doped $Bi_2Sr_2CuO_{6+\delta}$ single crystals are grown by the traveling solvent floating zone method [27]. Starting materials of $Bi_2O_3$, PbO, $Sr_2CO_3$ and CuO with 99.99% purity are mixed in an agate mortar and calcined at 750 ℃ - 810 ℃ in muffle furnace for 24 h. After pressed into a cylindrical rod in a hydrostatic pressure of ~ 70 MPa, the sample is sintered in a vertical molisili furnace at 840 ℃ for 48 h in air. The sintered rod was then pre-melted in the floating zone furnace at a traveling velocity of 25 ~ 30 mm/h to obtain a dense feed rod. By trying and optimizing the growing conditions, the dense feed rod was again melted on the seed rod just like the pre-melting process but with a slower travelling velocity of 0.5 mm/h. Single crystals of ~ 50 × 5 × 2 mm$^3$ can be obtained by cutting from the as-grown ingot.

**STM measurement.** The STM experiments are performed with an ultrahigh vacuum (UHV) and low temperature STM system. The overdoped Bi-2201 samples are cleaved in the UHV chamber at liquid nitrogen temperature (~77 K), and immediately inserted into the measurement stage. All STM results in this paper are acquired at liquid helium temperature (~4.5 K). The STM topography is typically taken with a sample bias $V = 100$ mV and a setpoint current $I = 100$ pA. The $dI/dV$ spectra are taken with a standard lock-in technique with modulation frequency $f = 983.4$ Hz and amplitude $V_{ac} = 3$ mV. The tips are fabricated from thin Tungsten wires (0.25-0.5 mm) by electrochemical reaction recipes, and treated by e-beam sputtering and field emission cleaning on the Au (111) crystal sample.

**Data availability.** The datasets generated during and/or analysed during the current study are available from the corresponding author on reasonable request.

**Reference**

1. Fradkin, E., Kivelson, S. A., Lawler, M. J., Eisenstein, J. P. & Mackenzie, A. P. Nematic Fermi fluids in condensed matter physics. *Annu. Rev. Condens. Matter Phys.* **1,** 153-178 (2010).

2. Keimer, B., Kivelson, S. A., Norman, M. R., Uchida, S. and Zaanen, J. From quantum matter to high-temperature superconductivity in copper oxides. *Nature* **518,** 179-186 (2015).

3. Chen, X. H., Dai, P. C., Feng, D. L., Xiang, T. & Zhang, F. C. Iron-based high transition temperature superconductors. *Natl. Sci. Rev.* **1,** 371-395 (2014).

4. Lawler, M. J. *et al.* Intra-unit-cell electronic nematicity of the high-$T_c$ copper-oxide pseudogap states. *Nature* **466,** 347-351 (2010).

5. Kaminski, A. *et al.* Spontaneous breaking of time-reversal symmetry in the pseudogap state of a high-$T_c$ superconductor. *Nature* **416,** 610-613 (2002).

6. Fauqué, B. *et al.* Magnetic order in the pseudogap phase of high-$T_c$ superconductors. *Phys. Rev. Lett.* **96,** 197001 (2006).

7. Li, Y. *et al.* Unusual magnetic order in the pseudogap region of the superconductor $HgBa_2CuO_{4+\delta}$. *Nature* **455,** 372-375 (2008).

8. Yi, M. *et al.* Symmetry-breaking orbital anisotropy observed for detwinned $Ba(Fe_{1-x}Co_x)_2As_2$ above the spin density wave transition. *PNAS* **108,** 6878–6883 (2011).

9. Mesaros, A. *et al.* Topological defects coupling smectic modulations to intra–unit-cell nematicity in cuprates. *Science* **333,** 426-430 (2011).

10. Rosenthal, E. P. *et al.* Visualization of electron nematicity and unidirectional antiferroic fluctuations at high temperatures in NaFeAs. *Nat. Phys.* **10,** 225-232 (2014).

11. Chu, J. H., Kuo, H. H., Analytis, J. G. & Fisher, I. R. Divergent nematic susceptibility in an iron arsenide susperconductor. *Science* **337,** 710-712 (2012).

12. Fujita, K. *et al.* Simultaneous transitions in cuprate momentum-space topology and electronic symmetry breaking. *Science* **344,** 612-616 (2014).

13. Kohsaka, Y. *et al.* How Cooper pairs vanish approaching the Mott insulator in $Bi_2Sr_2CaCu_2O_{8+\delta}$. *Nature* **454,** 1072-1078 (2008).

14. He, Y. *et al.* Fermi surface and pseudogap evolution in a cuprate superconductor. *Science* **344,** 608-611 (2014).

15. Boyer, M. C. *et al.* Imaging the two gaps of the high-temperature superconductor $Bi_2Sr_2CuO_{6+x}$. *Nat. Phys.* **3,** 802-806 (2007).

16. Slezak, J. A. *et al.* Imaging the impact on cuprate superconductivity of varying the interatomic distances within individual crystal unit cells. *PNAS* **105,** 3203-3208 (2008).

17. Hoffman, J. E. *et al.* Imaging quasiparticle interference in $Bi_2Sr_2CaCu_2O_{8+\delta}$. *Science* **297,** 1148-1151 (2002).

18. Wang, Q.-H. & Lee, D.-H. Quasiparticle scattering interference in high-temperature superconductors. *Phys. Rev. B* **67,** 020511(R) (2003).

19. Piriou, A., Jenkins, N., Berthod, C., Maggio-Aprile, I. & Fischer, Ø. First direct observation of the

Van Hove singularity in the tunnelling spectra of cuprates. *Nat. Commun.* **2,** 1229 (2011).

20. Hanaguri, T. *et al.* Quasiparticle interference and superconducting gap in $Ca_{2-x}Na_xCuO_2Cl_2$. *Nat. Phys.* **3,** 865-871 (2007).

21. Zeljkovic, I. *et al.* Scanning tunnelling microscopy imaging of symmetry-breaking structural distortion in the bismuth-based cuprate superconductors. *Nat. Mater.* **11,** 585-589 (2012).

22. Fujita, K. *et al.* Direct phase-sensitive identification of a *d*-form factor density wave in underdoped cuprates. *PNAS* **111,** E3026–E3032 (2014).

23. Hamidian, M. H. *et al.* Atomic-scale electronic structure of the cuprate *d*-symmetry form factor density wave state. *Nat. Phys.* **12,** 150-156 (2016).

24. Torre, E. G. D., He, Y. & Demler, E. Holographic maps of quasiparticle interference. *Nat. Phys.* **12,** 1052-1056 (2016).

25. Silva Neto, E. H. da *et al.* Ubiquitous interplay between charge ordering and high-temperature superconductivity in cuprates. *Science* **343,** 393-396 (2014).

26. Cai, P. *et al.* Visualizing the evolution from the Mott insulator to a charge-ordered insulator in lightly doped cuprates. *Nat. Phys.* **12,** 1047-1051 (2016).

27. Zhao, L. *et al.* High-quality large-sized single crystals of Pb-doped $Bi_2Sr_2CuO_{6+\delta}$ high-$T_c$ superconductors grown with traveling solvent floating zone method. *Chin. Phys. Lett.* **27,** 087401 (2010).

**Acknowledgements**

We thank F. C. Zhang for helpful discussions. This work is supported by the National Basic Research Program of China (2014CB921203, 2015CB921004), the National Natural Science Foundation of China (NSFC-11374260), and the Fundamental Research Funds for the Central Universities in China. XJZ thanks financial support from the NSFC (11190022 and 11334010), and the Strategic Priority Research Program (B) of CAS with Grant No. XDB07020300.

**Author contributions**

Y. Z. analyzed the data. Y. F. and K. L. B. performed the STM measurements. W. H. Z. participated in the measurements. Y. D. and X. J. Z. grew the single crystal samples. J. E. H. discussed the results and modified the manuscript. Y. Y. designed the project and prepared the manuscript. All authors have read and approved the final version of the manuscript.

**Additional Information**

**Supplementary information** accompanies this at http://www.nature.com/srep

**Competing financial interests:** The authors declare no competing financial interests.

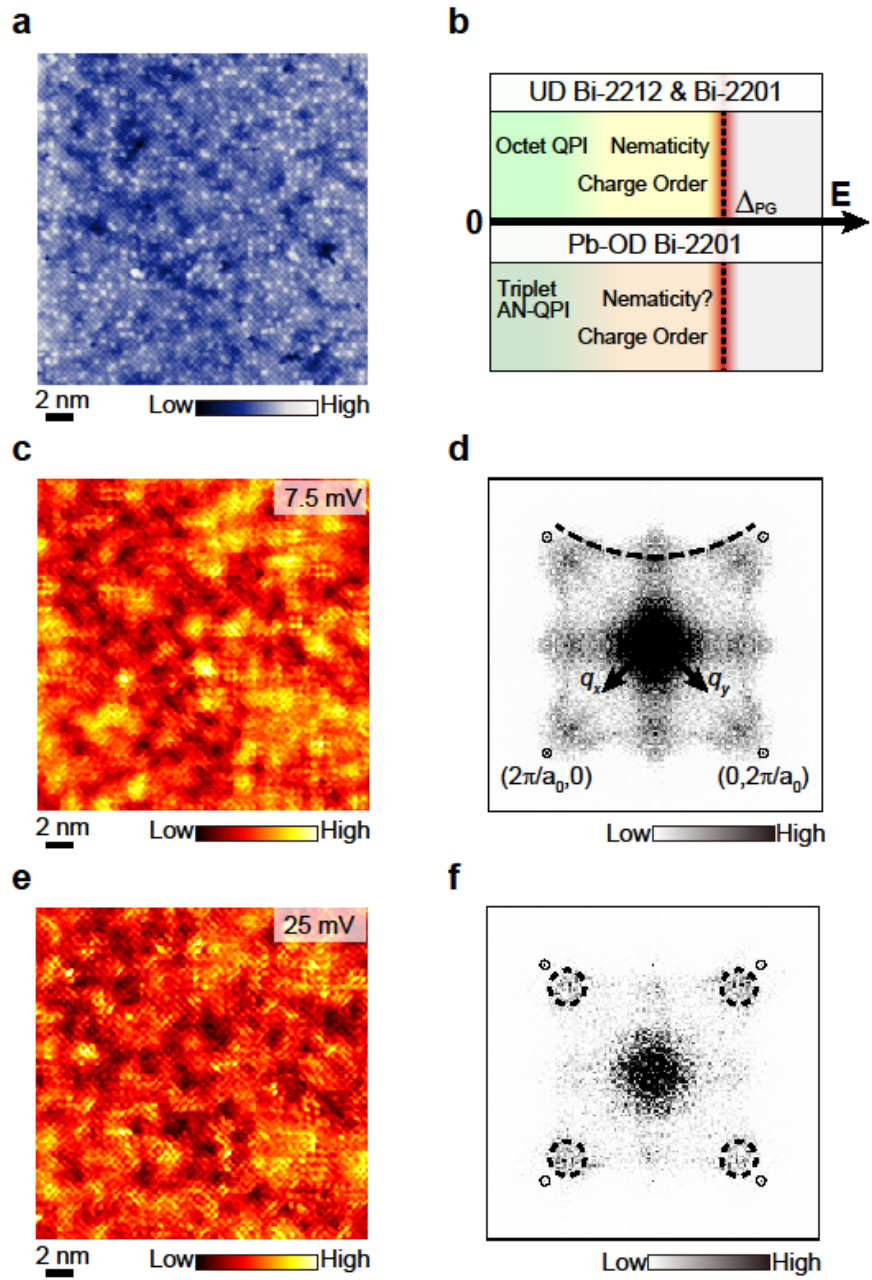

**Figure 1 | Topography and electronic properties of an overdoped Bi-2201 ($T_c = 13$ K).** (a) Topographical image ($270 \times 270$ Å$^2$) of a cleaved BiO layer with the bias voltage $V = 100$ mV and tunneling current $I = 100$ pA. The brighter spots in the lattice correspond to the Pb substitutes. (b) A diagram of different electronic behaviors, evolving as the energy increases from the Fermi energy (zero) to the PG energy $\Delta_{PG}$. (c) A differential tunneling conductance $dI/dV$ map measured at $V = 7.5$ mV on the same FOV as in **a**. (d) The $dI/dV$ map in the momentum space after the Fourier transform of panel **b**. The four Bragg peaks are labelled in the small circles. The dashed arc represents

a large FS. (**e**) Another $dI/dV$ map measured at $V = 25$ mV on the same FOV as in **a**. (**f**) The $dI/dV$ map in the momentum space after the Fourier transform of panel **e**. The four peaks representing a charge order are enclosed in the dashed circles inside the first Brillouin zone near Bragg peaks.

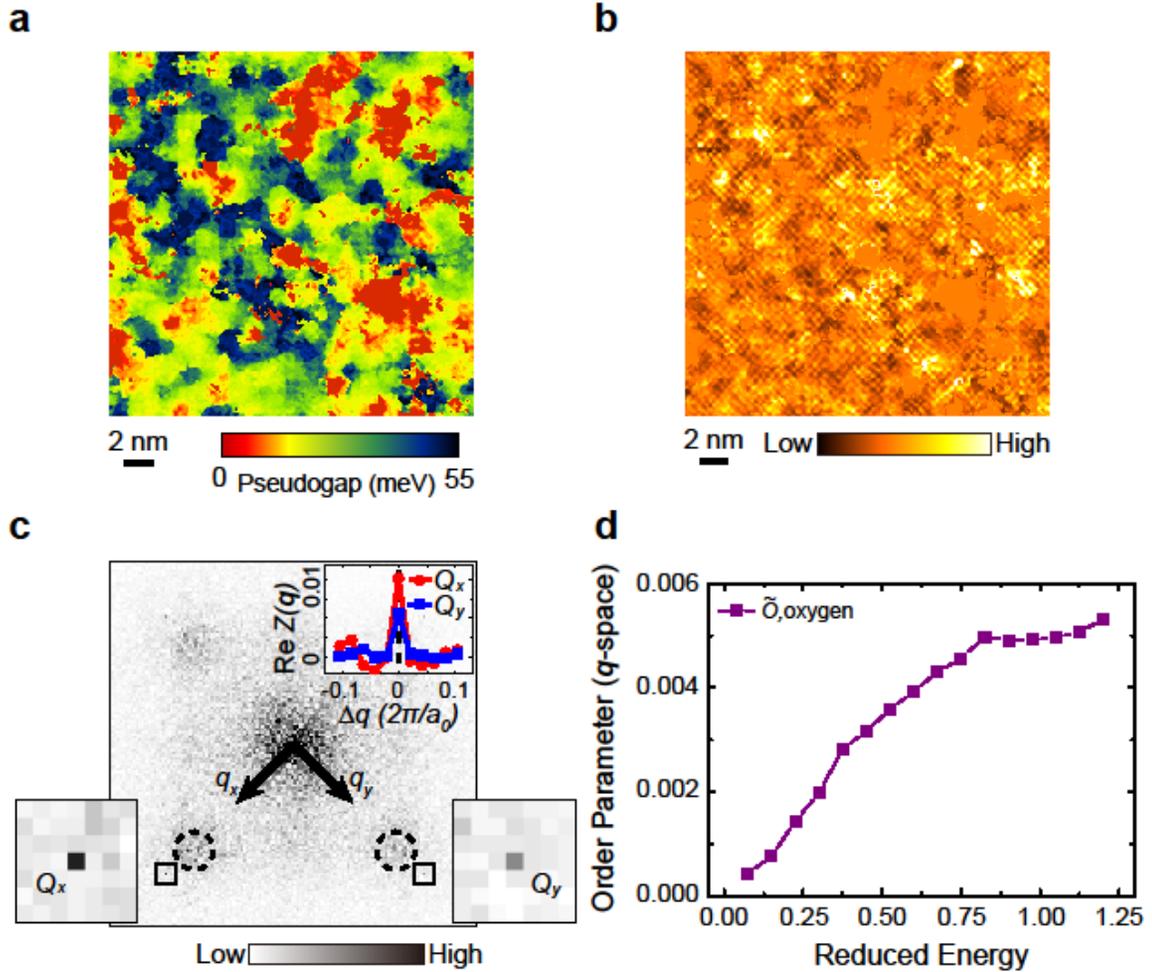

**Figure 2 | Ratio Z-map and collective nematic order ($q$-space).** (**a**) The PG map in the same FOV as in Fig. 1a. (**b**) A typical $Z(\vec{r}, \varepsilon)$ map at the reduced energy of $\varepsilon = 0.975$, with the VHS regions filled with the spatial average value. (**c**) The Fourier transform $\tilde{Z}(\vec{q}, \varepsilon)$ map of panel **b** in the momentum space. For the four Bragg peaks, the positions at $Q_x$ and $Q_y$ are highlighted and enlarged, in which both Bragg peaks collapse into one pixel. In the inset, the real parts of the line cuts at $Q_x$ and $Q_y$ are shown in red and blue colors, respectively. The charge order is prominent at different wave vectors, as labeled by the dashed circles. (**d**) The collective order parameter as a function of $\varepsilon$ in the momentum space.

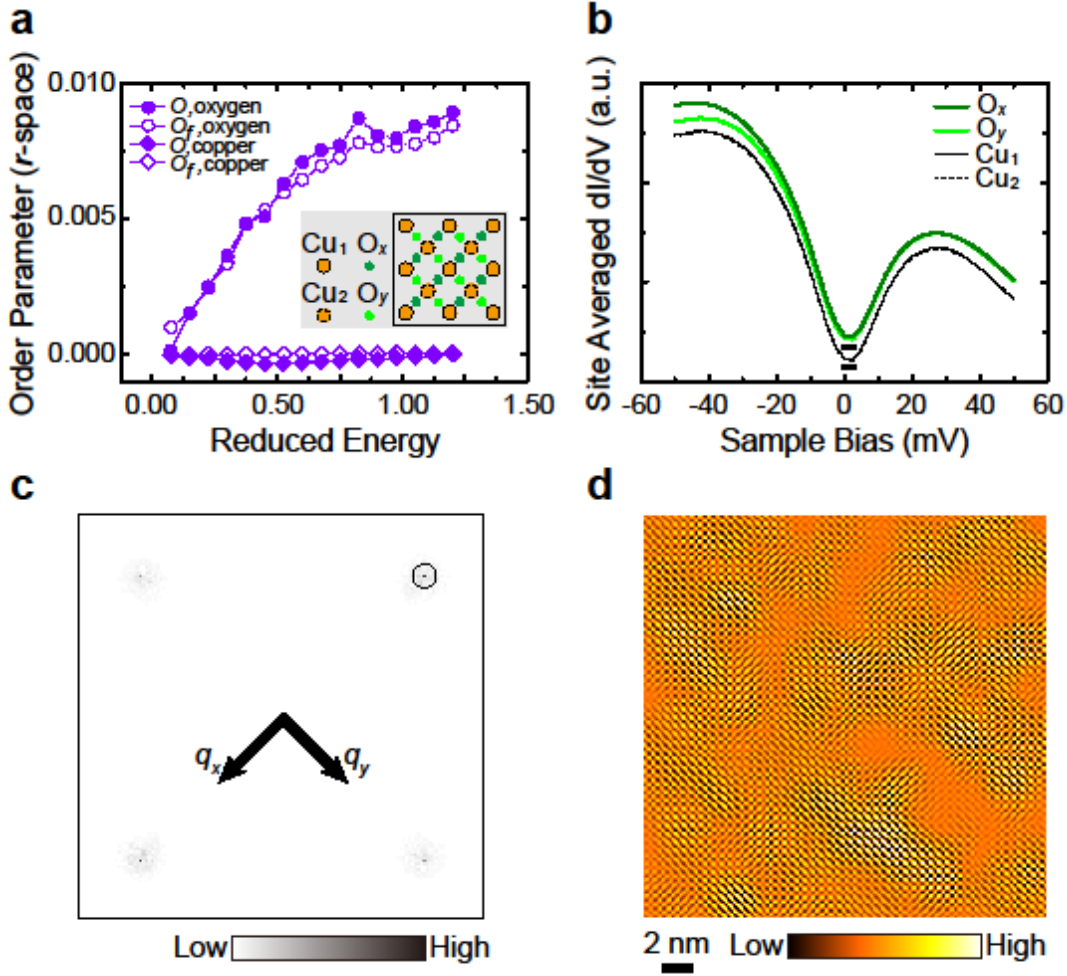

**Figure 3 | Filtered Z-map and collective nematic order (*r*-space).** (**a**) The collective order parameter in the real space as a function of $\varepsilon$. The lines with filled circles and diamonds are the results of $O(\varepsilon)$ from the O and Cu sites, respectively. The lines with open circles and diamonds are the results of $O_f(\varepsilon)$ from the same two types of sites after the filtration. The inset is a schematic of $CuO_2$ layer. (**b**) The averaged spectra on four different type of atomic sites ($O_x$, $O_y$, $Cu_1$, and $Cu_2$ as labeled in the inset of panel **a**.). The spectra for O sites are offset vertically for clarity. (**c**) The filtered ratio $\tilde{Z}_f(\vec{q}, \varepsilon)$ map in the momentum space after a Gaussian filtration around the four Bragg peaks. A circle centered at $Q_x$ with the radius $\Lambda^{-1} = 1.05$ nm is drawn to show the filtering size. (**d**) The filtered ratio $Z_f(\vec{r}, \varepsilon)$ map after the inverse Fourier transform of panel **c**.

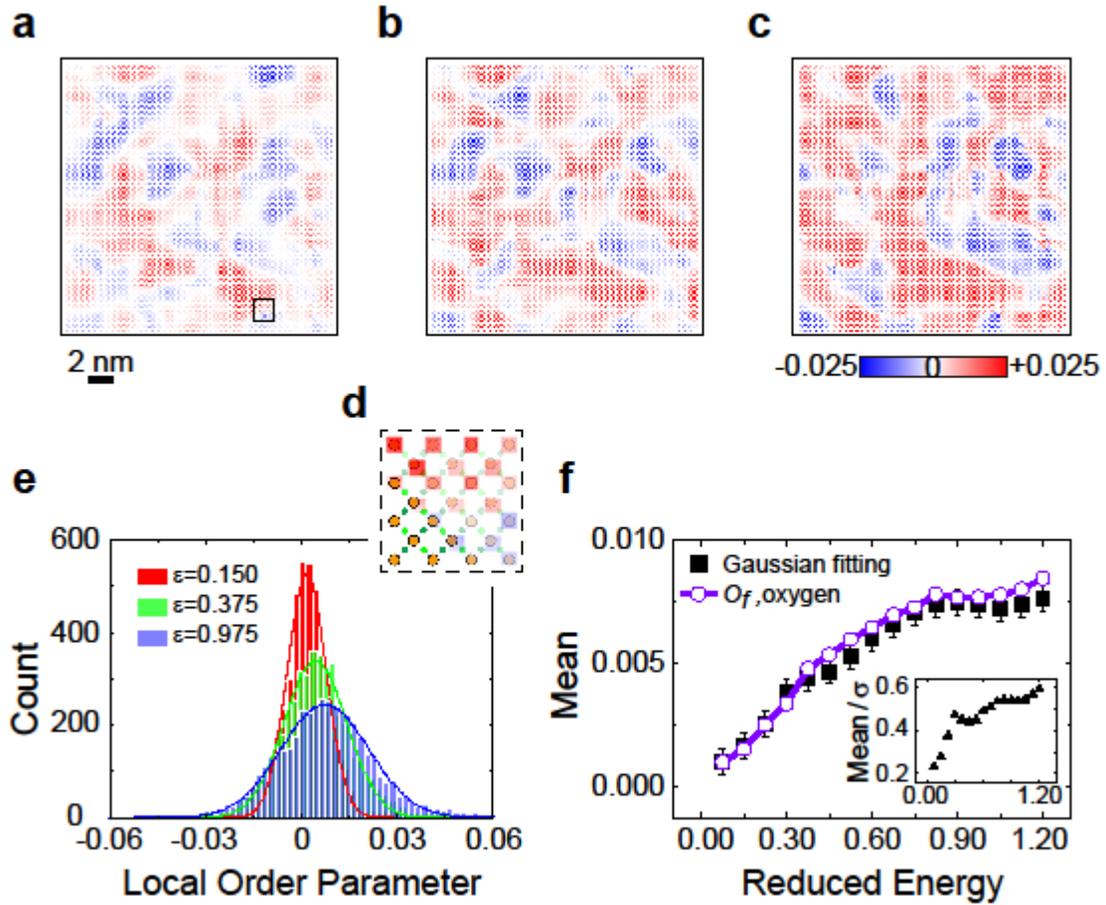

**Figure 4 | Distribution of Local nematic orders.** Maps of cell order parameters $O_f(n, \varepsilon)$ in the real space with the reduced energies of (**a**) $\varepsilon = 0.150$, (**b**) $\varepsilon = 0.375$, and (**c**) $\varepsilon = 0.975$. (**d**) The magnification of the selected region in **a**. The positions of Cu and O atoms are shown in the left corner, while the cell order parameters of unit cells are retained in the rest part. (**e**) Histograms of the cell order parameters from panels **a**-**c** with the bin size $\delta O_f = 0.001$. Each histogram is fitted by a dashed line of the Gaussian distribution. (**f**) The mean (filled squares) and ratio between the mean and relative standard deviation (filled triangular in the inset) of the Gaussian fitting as the function of $\varepsilon$. The error bar due to the bin size is shown for each mean of the Gaussian fitting. The collective order parameters $O_f(\varepsilon)$ (open circles) in Fig. 2e are provided for comparison.

**Supplementary Information for "The study of electronic nematicity in an overdoped (Bi, Pb)$_2$Sr$_2$CuO$_{6+\delta}$ superconductor using scanning tunneling spectroscopy"**

Yuan Zheng[1], Ying Fei[1], Kunliang Bu[1], Wenhao Zhang[1], Ying Ding[2], Xingjiang Zhou[2,3], Jennifer E. Hoffman[4], Yi Yin[1,5,*]

6. Department of Physics, Zhejiang University, Hangzhou, 310027, China
7. Beijing National Laboratory for Condensed Matter Physics, Institute of Physics, Academy of Science, Beijing 100190, China
8. Collaborative Innovation Center of Quantum Matter, Beijing 100871, China
9. Department of Physics, Harvard University, 17 Oxford St. Cambridge, 02138, USA
10. Collaborative Innovation Center of Advanced Microstructures, Nanjing 210093, China

    *Correspondence and requests for materials should be addressed to Y. Y. (email: yiyin@zju.edu.cn).

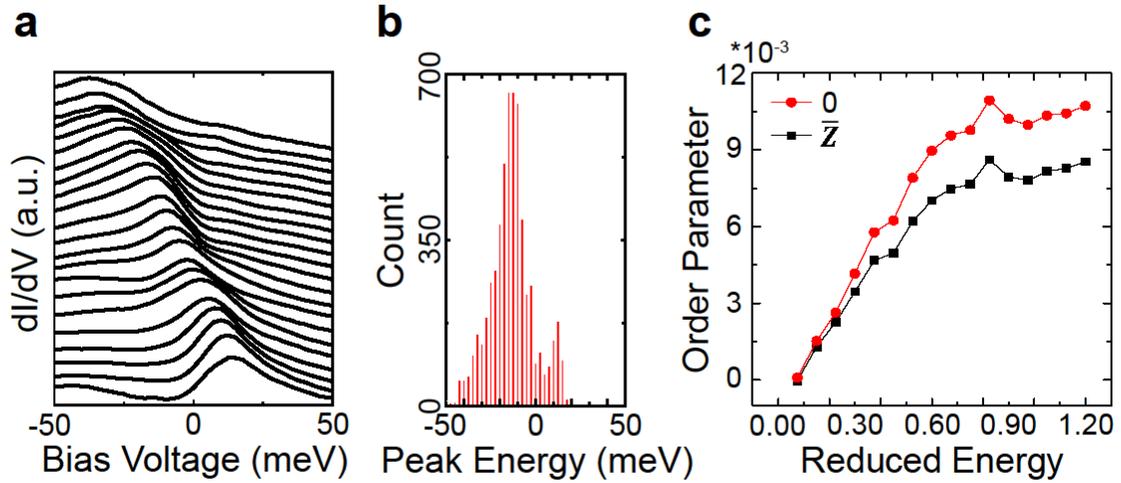

**Supplementary Figure S1 | Analysis on van Hove singularity region.** (**a**) The $dI/dV$ spectra with a single peak VHS in the same FOV as in the main text. Each curve is a spatial average of all spectra sharing the same peak position of $E_P$ binned by 2.5 meV. From top to bottom, the curves are sorted by increasing ranging from -37.5 meV to 15 meV, where a vertical offset is applied for clarity. (**b**) Histogram of the VHS peak energies. The average peak position is -10.5 meV and the standard deviation is 12.2 meV. (**c**) Reduced energy $\varepsilon$-dependence of the collective order parameters with the VHS region filled with zero (red curve) and $\bar{Z}$ (black curve).

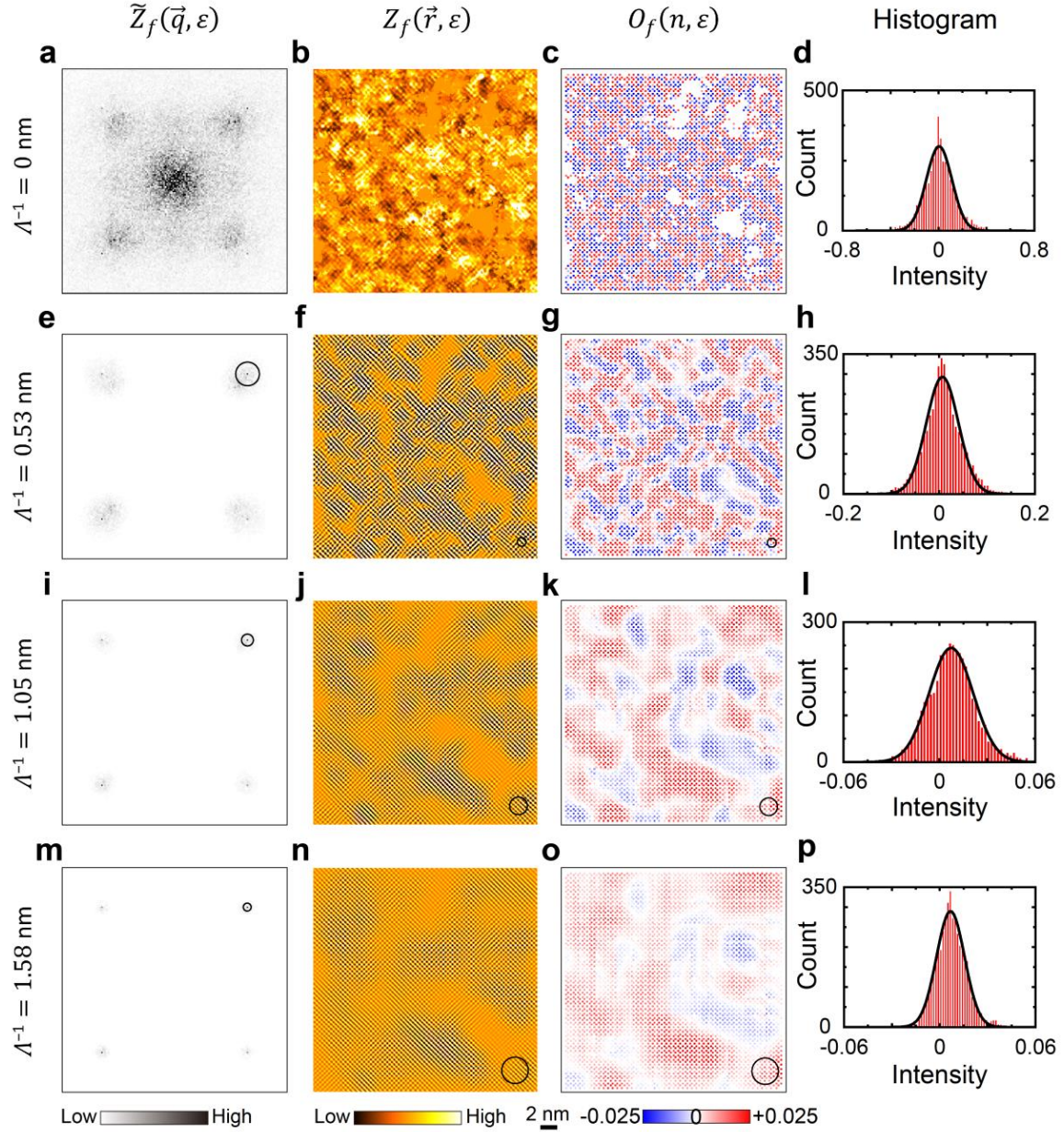

**Supplementary Figure S2 | Gaussian filtering procedure for $\varepsilon = 0.975$.** From left to right, figures (**a**, **e**, **i**, **m**) in the first column are the ratio maps ($\tilde{Z}(\vec{r}, \varepsilon)$ and $\tilde{Z}_f(\vec{r}, \varepsilon)$) in the momentum space; figures (**b**, **f**, **j**, **n**) in the second column are the ratio maps ($Z(\vec{r}, \varepsilon)$ and $Z_f(\vec{r}, \varepsilon)$) in the real space; figures (**c**, **g**, **k**, **o**) in the third column are the local nematic order maps ($O(n, \varepsilon)$ and $O_f(n, \varepsilon)$); figures (**d**, **f**, **j**, **n**) in the fourth column are the histogram of the corresponding cell nematic order parameters. From top to bottom, (**a**, **b**, **c**, **d**) in the first row are the unfiltered results; (**e**, **f**, **g**, **h**) in the second row are the filtered results with $\Lambda^{-1} = 0.527$ nm; (**i**, **j**, **k**, **l**) in the third row are the filtered results with $\Lambda^{-1} = 1.05$ nm; (**m**, **n**, **o**, **p**) in the fourth row are the filtered results with $\Lambda^{-1} = 1.58$

nm. The filtration size is shown by a black circle in each filtered map. For clarity of comparison, the unit cells with $O_f(n,\varepsilon) > 0.025$ and $O_f(n,\varepsilon) < 0.025$ are depicted in pure red and blue colors with the largest contrasts, respectively. The Gaussian fitting function for each histogram is represented by a solid curve.

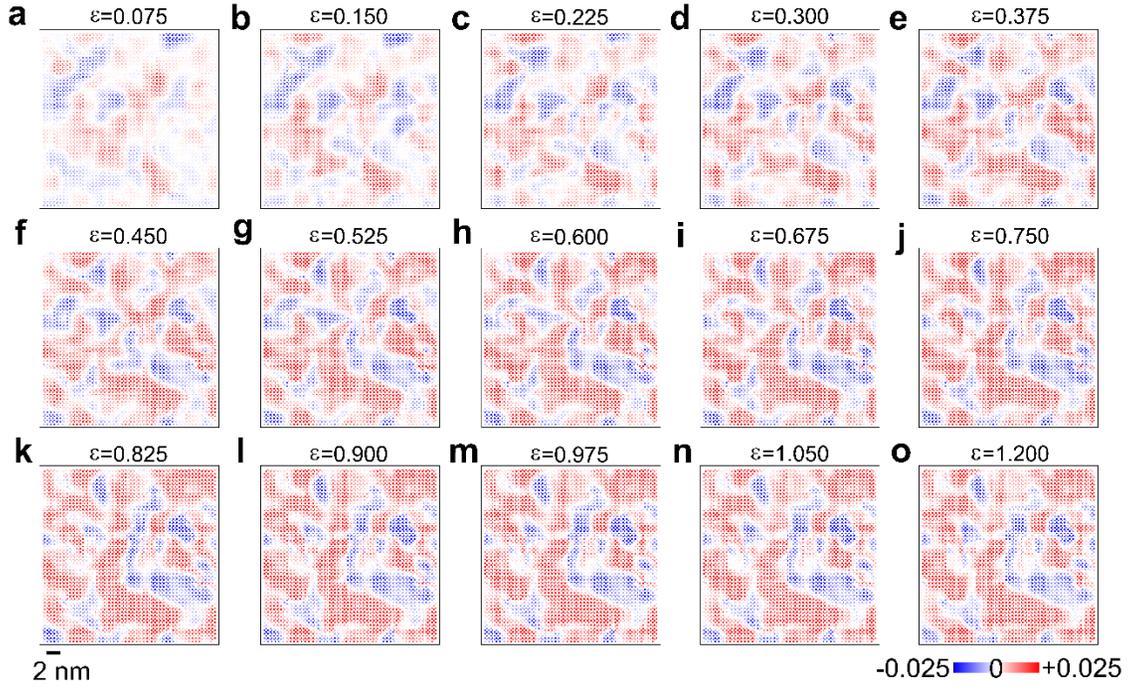

**Supplementary Figure S3 | Establishment of the nematic order.** (**a**)-(**o**) The evolution of the cell nematic order $O_f(n,\varepsilon)$ maps with the increase of the reduced energy from 0.075 to 1.200. All these maps are obtained under the Gaussian filtering procedure with the filtration size $\Lambda^{-1} = 1.05$ nm. For each reduced energy, the value of $\varepsilon$ is provided for the corresponding figure. For clarity of comparison, the unit cells with $O_f(n,\varepsilon) > 0.025$ and $O_f(n,\varepsilon) < 0.025$ are depicted in pure red and blue colors with the largest contrasts, respectively.

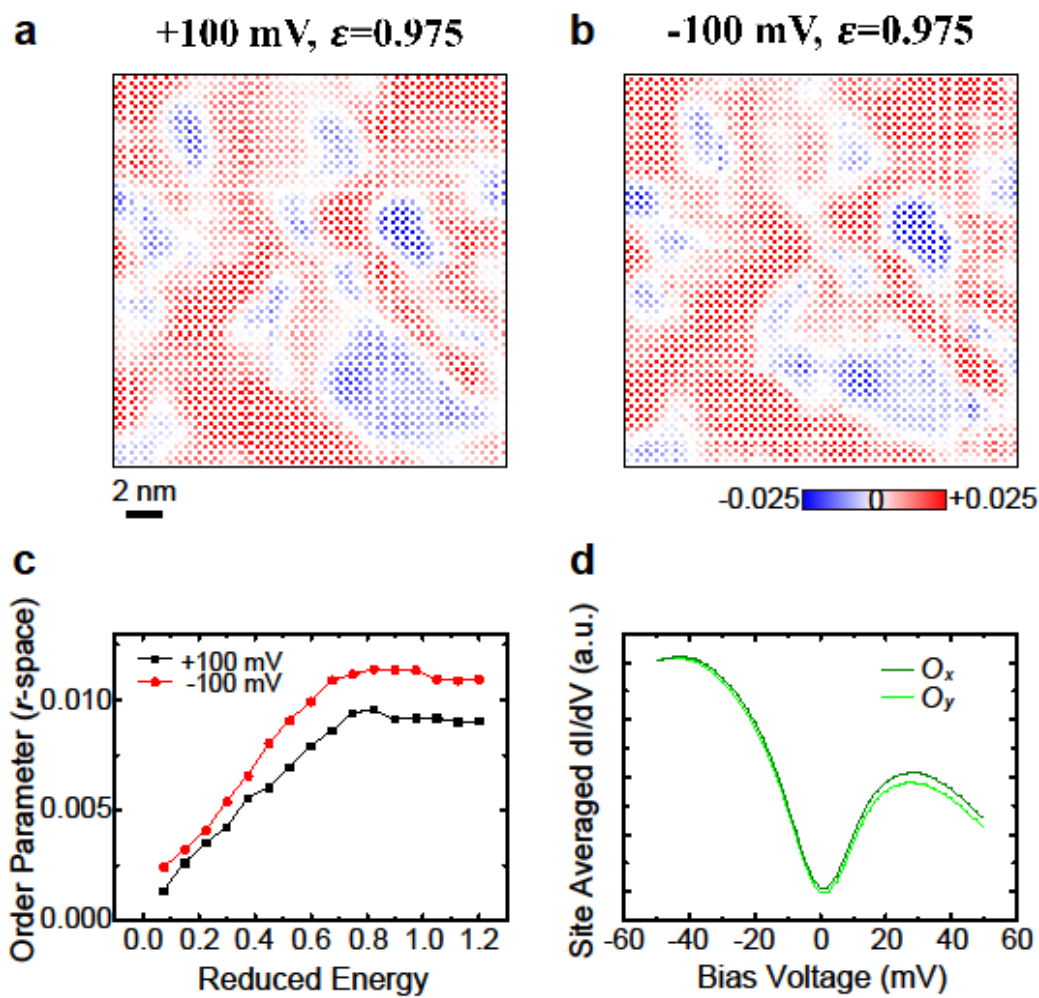

**Supplementary Figure S4 | The comparison of nematic order for two datasets with opposite sample bias polarities.** Maps of cell nematic order in the real space with reduced energy $\varepsilon = 0.975$, for two datasets with the sample bias voltage (**a**) $V_b = +100$ mV, and (**b**) $V_b = -100$ mV. (**c**) Reduced energy dependence of collective order parameters for both sample bias polarities. (**d**) The averaged spectra on $O_x$ and $O_y$ sites with the sample bias voltage $V_b = -100$ mV. With a negative sample bias, the spectral weight is normalized in the filled states (negative bias voltages), and the spectral shift mainly appears in the empty states (positive bias voltages).

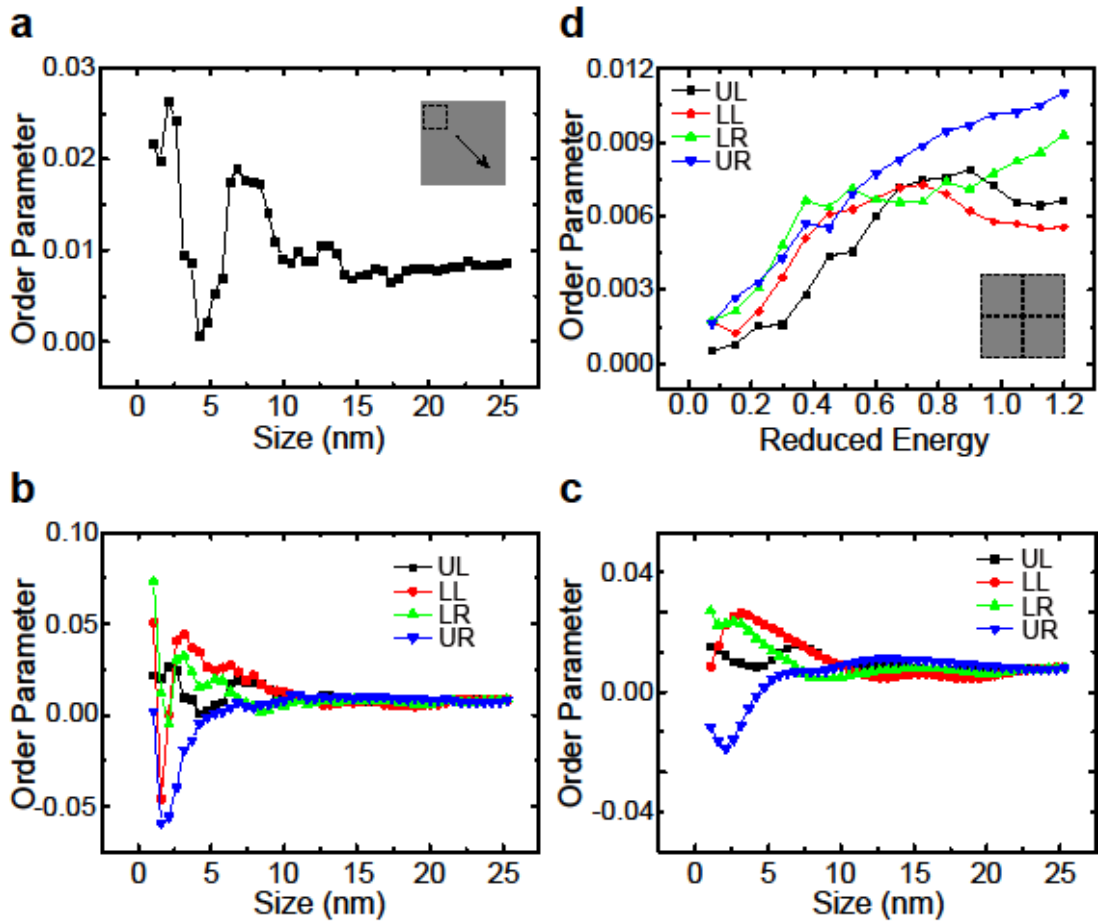

**Supplementary Figure S5 | Size effect of the collective nematic order.** (**a**) Size (side length) dependence of the collective order parameter of a square area. One corner of the square is fixed at the upper-left point of the same FOV as in the main text. (**b**) Size dependence of the collective nematic order extracted from the ratio map $Z(\vec{r}, \varepsilon = 0.975)$, for four different series of square areas and with corners fixed at four different points. UL, upper-left, LL, lower-left, LR, lower-right, UR, upper-right. (**c**) Size dependence of the collective nematic order similar as in panel **b**, but extracted from the filtered ratio map $Z_f(\vec{r}, \varepsilon)$. (**d**) Reduced energy dependence of the collective nematic order for four square areas at different locations, each with a size of $15 \times 15$ nm² (as schemed in the inset).

**Supplementary Note S1 | Differential conductance spectra with a van Hove singularity**

For the overdoped Bi-2201 sample studied in this paper, a pseudogap (PG) state can be determined for most of the differential conductance ($dI/dV$) spectra taken in the scanning field of view (FOV) (shown in Fig. 1a of the main text). Instead, a single logarithmic peak is observed around the Fermi energy level for a small portion (~10.9%) of $dI/dV$ spectra. Such a peak is considered as the signature of a van Hove singularity (VHS) in the electronic density of state (DOS). In Fig. S1a, a series of spatially averaged VHS spectra under a bin size of $\delta E_P = 2.5$ meV are presented as the VHS peak energy $E_P$ varies from -37.5 meV to 15 meV. As shown by the histogram in Fig. S1b, the majority of the VHS peak energies follows a distribution centered at -15 meV. However, hundreds of VHS peaks are extended to be above the Fermi energy level. In a previous study [18], the VHS spectra was found to appear as a non-cation-doped Bi-2201 sample approaches to the overdoped region. The positive PG peak is suppressed, while the original negative coherence peak is shifted to the Fermi energy level, gradually becoming a sharp VHS peak. In the whole FOV, the VHS areas are found to be surrounded by the PG areas with small gaps, and a smooth transition occurs between two characteristic spectra.

To minimize the influence of the VHS state in our analysis of the nematic order, the ratio $Z(\vec{r}, \varepsilon) = g(\vec{r}, \varepsilon)/g(\vec{r}, -\varepsilon)$ for positions inside the VHS region is replaced by a mean value, $\bar{Z}(\varepsilon)$. To make sure this assumption will not affect the main conclusion of nematic order in this paper, we also assign the ratio values in the VHS region with different constants, and compare the correspondingly analyzed results. In Fig. S1c, we show the collective order as a function of the reduced energy, with the VHS region set as zero and $\bar{Z}(\varepsilon)$ respectively. There is a similar trend of how the nematicity develops. The nematicity is qualitatively not affected by the artificial treatment of VHS region in our analysis. On the other hand, although a portion of ~10.9% VHS area is discovered in the FOV analyzed in this paper, this ratio is not a general percentage of VHS region in the current sample. The detailed study of the VHS region is not explored in this paper.

**Supplementary Note S2 | The Gaussian filtration procedure**

A key step of analyzing the nematic order in this paper is to apply a Gaussian filtering to the ratio $Z$-map in the momentum space. In Fig. S2, we take a typical example around the PG energy level, $\varepsilon = 0.975$, to demonstrate this filtration procedure. Figures S2a and S2b present the original $\tilde{Z}(\vec{q}, \varepsilon)$ and $Z(\vec{r}, \varepsilon)$ maps, which include information of all the wavelengths in addition to the intra-unit-cell modulation. A cell nematic order, $O(n, \varepsilon) = [Z_y(\vec{r}_n, \varepsilon) - Z_x(\vec{r}_n, \varepsilon)]/\bar{Z}(\varepsilon)$, is subsequently defined for each $n$th unit cell in the real space. The resulting $O$-map is displayed in Fig. S2c, from which we observe a significant inhomogeneity of both positive and negative nematicities. The spatial average, $O(\varepsilon) = \sum_n O(n, \varepsilon)/N$, leads to a collective order, $O(\varepsilon = 0.975) = 0.008$. The histogram of $O(n, \varepsilon)$ with the bin size $\delta O = 0.02$ in Fig. S2d is fitted by a Gaussian distribution with a mean $O^G(\varepsilon = 0.975) = 0.0052$ and a standard deviation $\sigma(\varepsilon = 0.975) = 0.107$. The atomic-site-specified nematic order is hard to be extracted from this real space distribution due to a large deviation between $O(\varepsilon)$ and $O^G(\varepsilon)$ as well as an extremely strong background noise, $\sigma(\varepsilon)/O^G(\varepsilon) \approx 20$.

Since the intra-unit-cell modulation is reflected by the Bragg peaks in the $\tilde{Z}(\vec{q}, \varepsilon)$ map, we can concentrate on signals near these peaks by filtering out interference information from other spatial modulations. In practice, a smooth Gaussian cut off, $f_\Lambda(\vec{q}) = \exp(-q^2/2\Lambda^2)$, is applied to each Bragg peak, which gives rise to a filtered signal,

$$\tilde{Z}_f(\vec{q}, \varepsilon) = \tilde{Z}(\vec{q}, \varepsilon)[f_\Lambda(\vec{q} + \vec{Q}_x) + f_\Lambda(\vec{q} - \vec{Q}_x) + f_\Lambda(\vec{q} + \vec{Q}_y) + f_\Lambda(\vec{q} - \vec{Q}_y)].$$

Figure S2e presents a filtered $\tilde{Z}_f(\vec{q}, \varepsilon)$ map under the filtration size of $\Lambda^{-1} = 0.527$ nm, which is equal to 5 pixels of the FOV (the total $256 \times 256$ pixels). Next we can take an inverse Fourier transform of $\tilde{Z}_f(\vec{q}, \varepsilon)$ to obtain the filtered ratio map, $Z_f(\vec{r}, \varepsilon)$, in the real space. Compared to the unfiltered map in Fig. S2b, $Z_f(\vec{r}, \varepsilon)$ in Fig. S2f exhibits lattice-like patterns since a locally weighted average is considered for each spatial position. Accordingly, the filtered $O_f(n, \varepsilon)$ map in Fig. S2g is less fluctuated than the unfiltered map in Figure S2c, and inhomogeneous domains of the two opposite nematicities are observed. The parameters of the Gaussian fitting function for the histogram in Fig. S2h are $O_f^G(\varepsilon = 0.975) = 0.0067$ and $\sigma_f(\varepsilon = 0.975) = 0.034$. Although the mean value $O_f^G(\varepsilon)$ is closer to the collective nematic order, $O_f(\varepsilon = 0.975) = 0.0079$, the background noise still dominates with $\sigma_f(\varepsilon)/O_f^G(\varepsilon) \approx 5$.

Next we consider a larger filtration size, $\Lambda^{-1} = 1.05$ nm, which is equal to 10 pixels of the FOV. The maps of $\tilde{Z}_f(\vec{q}, \varepsilon)$, $Z_f(\vec{r}, \varepsilon)$, and $O_f(n, \varepsilon)$ following the filtering procedure are plotted in Fig. S2i, S2j, and S2k, respectively. The lattice-like patterns in $Z_f(\vec{r}, \varepsilon)$ and the real-space nematic domains in $O_f(n, \varepsilon)$ are further strengthened although more information is filtered out at the same time. For the histogram in Figure S2l, the parameters of the Gaussian fitting function are $O_f^G(\varepsilon = 0.975) = 0.0074$ and $\sigma_f(\varepsilon = 0.975) = 0.014$. The mean value is nearly the same as the collective order, $O_f(\varepsilon = 0.975) = 0.0077$, and the nematic peak can be distinguished from the background noise with $\sigma_f(\varepsilon)/O_f^G(\varepsilon) \approx 2$.

The third filtration size studied is $\Lambda^{-1} = 1.58$ nm, which is equal to 15 pixels of the FOV. The maps of $\tilde{Z}_f(\vec{q}, \varepsilon)$, $Z_f(\vec{r}, \varepsilon)$, and $O_f(n, \varepsilon)$ following the filtering procedure are plotted in Figures S2m, S2n, and S2o, respectively. As the contrast of the $Z_f(\vec{r}, \varepsilon)$ map decreases significantly, the fluctuations of local nematic domains in the $O_f(n, \varepsilon)$ map are further suppressed. For the histogram in Figure S2p, the parameters of the Gaussian fitting function are $O_f^G(\varepsilon = 0.975) = 0.0068$ and $\sigma_f(\varepsilon = 0.975) = 0.088$. Although the nematic peak is further narrowed with $\sigma_f(\varepsilon)/O_f^G(\varepsilon) \approx 1.3$, it becomes more difficult for us to extract information of atomic-site-specified nematicity, which could be essential for understanding the mechanism of nematicity.

Therefore, an appropriate filtration size $\Lambda^{-1}$ is required in our analysis of the nematicity. On one hand, the filtered $Z_f(\vec{r}, \varepsilon)$ map is expected to be weakly correlated with the unfiltered $Z(\vec{r}, \varepsilon)$ map so that the dominant background interference signal is efficiently filtered out. On the other hand, $\Lambda^{-1}$ is required to be within the average size of the electronic order, e.g., the PG state, so that the locally weighted average is roughly taken within the area of the same $\Delta_{\text{PG}}$. In the main text, all the filtered results are calculated using $\Lambda^{-1} = 1.05$ nm, which can satisfy both requirements and is also close to the filtration size in a previous study of Bi-2212 [3]. However, we must emphasize that the qualitative conclusion of this paper is unaffected by the choice of $\Lambda^{-1}$. The establishment of the collective nematic order $O_f(\varepsilon)$ is consistently observed as $\Lambda^{-1}$ increases. In addition, a broad distribution of $O_f(n, \varepsilon)$ is sustained for a large filtration size, $\Lambda^{-1} = 1.58$ nm, which confirms the coexistence of the two opposite nematicities and their strong spatial inhomogeneity.

**Supplementary Note S3 | Dataset taken with a negative sample bias voltage**

As an intrinsic electronic property, the nematic order should not be dependent on the tunnel junction condition with which the spectra dataset is taken. In the same FOV as in the main text, we have changed the sample bias voltage to $V_b = -100$ mV and taken another dataset of $dI/dV$ spectra. The similar analysis has been made to explore the nematic order. In Fig. S4a and S4b, we compare the cell nematic order maps for $V_b = +100$ mV and $V_b = -100$ mV. The similar order patterns are displayed, with very slight difference in details. The collective orders as a function of reduced energy are also compared in Fig. S4c. Qualitatively, they show the similar trend of nematicity. Furthermore, the averaged spectra on the $O_x$ and $O_y$ sites are compared in Fig. S4d. With the negative sample bias condition, the spectral weight is normalized for filled states (negative bias voltages) and the spectral shift mainly appears in the empty states (positive bias voltages). We consider the two detected nematic order consistent with each other, although the datasets are taken one after another and with different tunnel junction conditions.

**Supplementary Note S4 | Size effect of the collective nematic order**

From the cell nematic order maps we analyze and display in the main text, there is a broad real-space distribution of the site-specified nematicity, and a real space fluctuation of positive and negative regions of nematicity. The real space order fluctuation is also reproducible as shown in the supplementary note S3. With this real space fluctuation, we should reconsider how the collective order evolves as the size of the analyzed area increases. For the FOV in the main text, an area of $27 \times 27$ nm$^2$, we define a series of square areas with different sides, but sharing the same upper-left corner. For each square area, the collective nematic order is correspondingly calculated for the ratio maps $Z(\vec{r}, \varepsilon)$. The size dependence of the collective order is displayed in Fig. S5a. We could notice that the collective nematic order oscillates a lot for the areas with the area smaller than $\sim 10 \times 10$ nm$^2$, depending on how the positive or negative regions of nematicity aggregate in the small area. But when the size increases, the collective nematic order converges to a positive value.

In Fig. S5b, we display the similar order evolution as the size increases for square areas starting from four different corners in the same FOV. The initial oscillations of the order for small square areas vary dramatically among four different series, illustrating the location dependence of evolved orders. The range of oscillation is much larger than the final converged order values, because we analyze the data based on unfiltered ratio maps $Z(\vec{r},\varepsilon)$. The range of initial oscillation is reasonably smaller when the filtered ratio maps $Z_f(\vec{r},\varepsilon)$ are applied to obtain the size dependence of nematic order evolution, as shown in Fig. S5c.

In Fig. S5d, we roughly separate the original FOV to four different areas, each with a size of $15 \times 15$ nm$^2$, and display their corresponding $\varepsilon$-dependence of collective nematic order. They share a similar trend of nematicity, although the four areas are at different locations. With the size-effect analysis of collective nematic order as above, we consider the FOV of $27 \times 27$ nm$^2$ is a reasonable large enough size to define the collective nematic order. Although the quantitative value of nematic order should vary with the location and size of the FOV, we claim a qualitative discovery and mapping of nematicity in the over-doped sample in this paper.